\documentstyle[epsf]{mn}

\newcommand{\be}{\begin{equation}}
\newcommand{\ee}{\end{equation}}
\newcommand{\msol}{$\mathrm{M}_{\sun}$}
\newcommand{\ud}{\mathrm{d}}

\begin{document}

\title[Wide binaries in the Orion Nebula Cluster]{Wide binaries in the
Orion Nebula Cluster}

\author[A. Scally, C. Clarke and M. J. McCaughrean]{Aylwyn Scally,$^{1}$
Cathie Clarke$^1$ and Mark J. McCaughrean$^2$\\
$^1$Institute of Astronomy, Madingley Road, Cambridge CB3 0HA, England\\
$^2$Astrophysikalisches Institut Potsdam, An der Sternwarte 16, D14482
Potsdam, Germany}

\maketitle

\begin{abstract}
Using proper motion data for 894 stars in the Orion Nebula Cluster
(ONC) compiled by Jones \& Walker in 1988, we search for binaries with
apparent separations in the range 1000--5000 AU, and find an upper
limit of three. Using a Monte Carlo method, we test the consistency of
this result with two hypotheses: i) that the cluster contains a binary
population identical to that found in the solar neighbourhood, and ii)
that the cluster contains no binaries at all in this separation range.
We obtain results strongly favouring the latter hypothesis.

Star formation in the Galaxy is seen to occur in a variety of
different environments, but it has been proposed that most stars may be
formed in dense regions similar to the ONC, rather than in less dense
groupings like that found in Taurus-Auriga. Since roughly 15 per cent
of galactic field stars are known to be in binaries with separations
greater than 1000~AU, the apparent absence of such binaries in the ONC
places an upper limit on the contribution that dense clusters can make
to galactic star formation. \end{abstract}

\begin{keywords}
stars: formation -- stars: pre-main-sequence -- stars: statistics --
open clusters and associations: individual: Orion Nebula Cluster --
binaries: general.
\end{keywords}

\section{Introduction}

Amongst the stars around us in the Galactic field, single systems like
our own are a minority. Most stars are found in binaries, or in systems
of even higher multiplicity. Recent surveys of the nearby stellar
population \cite{DM91,FM92}, extending to 20~pc or so, indicate that
about 60 per cent of systems in the solar neighbourhood are binary. In
younger populations, such as those in nearby star-forming regions, the
degree of binarity is often even higher. Surveys of the low-mass dark
cloud complexes in Taurus-Auriga, Ophiuchus, Chameleon, Lupus, and
Corona Australis reveal a fraction of binary systems roughly twice as
high as that of the solar neighbourhood in the range of binary
separations detected (typically 10--1500~AU) \cite{G+97,KLe98}. Thus
single stars are rare in these clusters. An exception to this trend is
the Orion Nebula Cluster (ONC) in the Orion~A giant molecular cloud,
which has a binary fraction roughly the same as the solar neighbourhood
value, at least over the separation range of $\sim$~25--500~AU
\cite{P+94,PaStG97,Pe+98,Pe98,SiClBe99}.

The ONC differs from the other regions mentioned in that it is densely
clustered, and contains a number of massive O and B type stars. In
particular, the core of the ONC, known as the Trapezium cluster after
the eponymous four central OB stars which dominate it \cite{HeTe86},
contains some 1000 stars with a peak density of 2--5$ \times 10^4$
stars $\mathrm{pc^{-3}}$ \cite{McS94,HHa98}. This compares to core
densities in the Taurus-Auriga complex of only around 10 stars
$\mathrm{pc^{-3}}$. It has been argued, based on estimates of star
formation rates \cite{MiSc78}, that dense OB associations like the ONC
are responsible for the majority of star formation in the Galaxy, and
are therefore representative of the typical environment for early
stellar evolution.

The results on binarity in star-forming regions would seem to support
this conclusion. However, the high density of the ONC has meant that
the surveys carried out there so far have only been able to detect
binaries separated by less than about 500 AU\@. It would be very useful
to examine the binary fraction at wider separations, but in doing so
one is increasingly likely to observe chance projected alignments
rather than actual bound pairs. Bate, Clarke \& McCaughrean
\shortcite{BCMc98} analysed the mean surface density of companions in
the centre of the ONC using data from Prosser et al. \shortcite{P+94}
and McCaughrean et al. \shortcite{Mc+96}, and found weak evidence
there for a deficit of binaries with separations greater than 500~AU.
But to do better, and to detect wider binaries properly, one must
use the fact that they have a common proper motion across the
sky. Specifically, one looks for stars whose relative velocity is
less than the critical value needed to escape their mutual
gravitational attraction (which can be determined by making an
assumption or estimate of the stars' masses).

In this work we attempt to detect common proper motion binaries in the ONC
using the data obtained and catalogued by Jones \& Walker (1988; hereafter
also JW), and thence determine whether the number found is consistent with
the field star binary distribution over the same separation range.

\section{Finding common proper motion pairs}

The JW study used a series of optical wavelength photographic plates
taken over a baseline of about 20 years, and obtained proper motion
measurements for over 1000 stars. Compared to deeper and higher spatial
resolution surveys carried out using the Hubble Space Telescope at
optical wavelengths \cite{P+94} and ground-based adaptive optics
techniques at infrared wavelengths \cite{McS94}, the JW data reveal
only a fraction of the total stellar population. However, it remains
the best proper motion survey to date, and covers a relatively large
area on the sky, extending roughly 15 arcmin or 2~pc from the cluster
centre, taken to be $\theta^{1}$~Ori~C, the most massive of the four
Trapezium stars.

Ideally, one would have data in \emph{three} dimensions for position
and velocity, plus the mass of each star, as it would then be a simple
task to identify all the bound pairs within the cluster, and give an
accurate value for the binary fraction. Instead, because the data
provides information only in two dimensions, one can do no more than
place an upper limit on the number of possible bound stars. The
approach taken here is to look for \emph{apparent binaries} in the JW
data -- pairs whose 2D positions and velocities allow them to be bound
if one ignores the third dimension -- and then to compare the number of
these with the number of similar pairs observed in a randomly generated
model cluster with known parameters.

Hillenbrand \shortcite{H97} found the mean stellar mass in the
Trapezium to be approximately 0.8 $\mathrm{M_{\sun}}$; here we make the
assumption that all the stars are of solar mass.
%This is a conservative
%assumption, as it leads to an overestimate of the number of bound pairs
%in the cluster.
The distance to the ONC is taken to be 470 pc (the value used by JW), at which
1 pc corresponds to 7.3 arcmin on the sky. We test for binarity at
projected separations up to 5000~AU, beyond which the critical relative
velocity (the escape velocity for a pair at that separation) is lower
than the mean error in the JW velocity data ($\sim
0.8\ \mathrm{km\,s^{-1}}$). A lower limit of 1000~AU is imposed by the
observational resolution of $\sim$ 2 arcsec. Thus our study is
complementary to previous work, where binaries were sought at
separations roughly ten times smaller.

An analysis of the JW data gives the following initial results.
Eliminating stars with less than 80 per cent membership probability
leaves 894 stars in the cluster catalogue, which give a total of $894
\times (894 - 1) / 2 = 399171$ pairs. Of these, 192 have apparent
separations between 1000 and 5000 AU\@. Assuming solar masses and using
the JW velocity measurements (ignoring errors for the time being),
three of these 192 have relative velocities lower than the critical
velocity for their separation.

%\vspace{12pt} \noindent
It is perhaps likely that if data for the radial dimension were known,
these binaries would turn out to be unbound. But it is also likely that
the effect of errors has been to `disrupt' other apparent binaries in
the cluster. These complications -- in particular the projection effect
-- mean that we cannot directly compare our result with a prediction
from the Duquennoy \& Mayor (or any other) period distribution.
Instead, as mentioned above, the comparison is made by constructing an
ensemble of model clusters whose observable parameters match the JW
data statistically, but which contain the population of binaries we
want to compare with.

\section{Simulated clusters}

\begin{figure}
\epsfxsize=8.0truecm\epsfbox{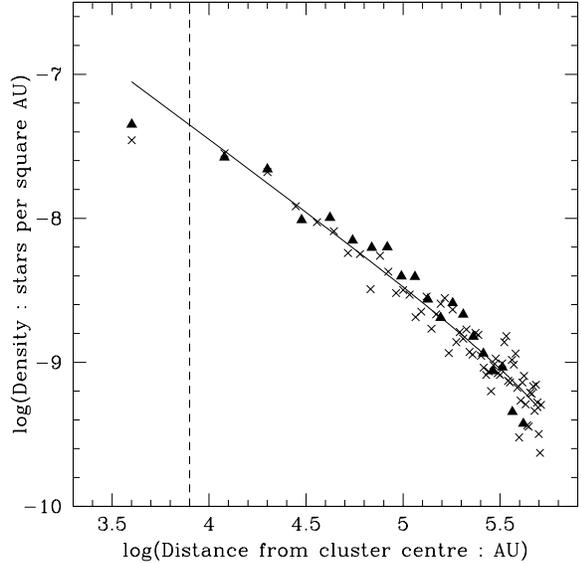}
\caption{The radial density distribution of a typical model cluster
($\times$ symbols) compared to that of the ONC as surveyed by Jones \&
Walker \shortcite{JW88} (triangles). The solid curve shows what
gradient would be expected for a singular isothermal sphere of radius 5~pc.
The effect of the finite boundary can be seen as a slight downturn with
increasing radius. Similarly, the effect of the flat core can be seen
in the data points as a flattening towards small $r$. The dashed line
marks the core radius.}
\label{F:radDist}
\end{figure}

The model clusters have a three-dimensional density distribution
corresponding to an isothermal sphere with a flat core:
\be
\rho(r) = \left\{ \begin{array}{ll} \rho_{0} & \textrm{if $r \le R_{\mathrm{core}}$} \\
\rho_{0} \left( \frac{R_{\mathrm{core}}}{r} \right)^2 & \textrm{if $r \ge R_{\mathrm{core}}$}
\end{array} \right. \label{E:truncIsotherm}
\ee
where $R_{\mathrm{core}} = 0.04\ \mathrm{pc}$. A similar distribution
was shown by Bate, Clarke, \& McCaughrean \shortcite{BCMc98} to give a
good fit to the projected density distribution of the JW data. Figure
\ref{F:radDist} shows a comparison between the simulated and observed
data.
A finite radius of 5~pc is chosen, within which a sufficient number of
stars are distributed according to (\ref{E:truncIsotherm}) to ensure
that, after the corrections discussed below, roughly 900 appear
projected within the same radius as the JW survey ($\sim 2$~pc). Each
system is then assigned a velocity chosen from a Maxwellian
distribution with dispersion of 4~$\mathrm{km\,s^{-1}}$, equal (in
three dimensions) to that found by Jones \& Walker for most of the
cluster.

Two different binary populations are considered, one with a binary
fraction of zero (i.e. no binaries at all), and the other with a
population matching that found by Duquennoy \& Mayor \shortcite{DM91}.
In the latter, the periods have a log-normal distribution peaking at
180 years, which corresponds to a typical separation for solar mass
stars of 30~AU\@.

Once a cluster has been generated with these parameters, we mimic the
process of observing it, applying the same errors and restrictions to
the data as were present in Jones \& Walker's survey. For example, the
velocity errors in the JW catalogue were found to be log-normally
distributed with a peak at 0.8~$\mathrm{km\,s^{-1}}$, and we apply
noise to the velocities in the model cluster using the same
distribution. Similarly, the JW survey had a resolution limit of about
2 arcsec or 1000 AU, so any model stars which appear closer in
projection are merged into one.

We must also correct for the fact that some of the binaries present in
Duquennoy \& Mayor's population would have appeared as single stars to
Jones \& Walker, because one companion would have been fainter than
they could detect. In the outer part of the cluster, JW give their
detection limit as $I \sim 16$, but in the centre, where there is
bright background emmission from the Orion Nebula, this limit is
reduced. Indeed, within about 2 arcmin (0.3 pc) of the centre they
apparently detect no stars fainter than $I = 14$. We take the mean
reddening of $A_V = 2.4$ magnitudes measured towards the
optically-visible stars by Herbig \& Terndrup \shortcite{HeTe86}, which
corresponds to a mean reddening in the $I$ band of 1.4 magnitudes.
Then, in the outer regions, the JW study is sensitive to stars with
intrinsic $I$ magnitudes as faint as 14.6, which at the assumed
distance and an age of 1~Myr (see \S 4) corresponds to a mass of
roughly 0.2~\msol\ \cite{DaMz94}. The brighter detection limit within
0.3~pc corresponds to a mass of 0.6~\msol.

The minimum companion mass in the wide binaries included in Duquennoy
\& Mayor's survey was also about 0.2~\msol, this limit being determined
by the magnitude range of the data they used -- see Patience et al.\
\shortcite{Pt+98}. We therfore choose to apply a correction only to
the centre of our simulated clusters,
removing a certain fraction of the binary components whose projected
distance from $\theta^{1}$~Ori~C is less than 0.3 pc.
Denoting this fraction by $X$, we need to evaluate the following
integral:
\be
X = \int_{0.6}^{\infty}\left(\frac{\int_{0.2}^{0.6}s(m, \mu)
\,\ud \mu}{\int_{0.2}^{m}s(m, \mu) \,\ud \mu}\right) f(m) \,{\ud} m
\ee
where $f(m)$ is the stellar mass function in the cluster, and $s(m,
\mu)$ is the companion mass function -- the distribution of companions
of mass $\mu$ to primaries of mass $m$. For simplicity, we assume that
the two mass functions match in the region where $\mu \le m$:
\be
s(m, \mu) = \left\{ \begin{array}{ll}
f(\mu) & \textrm{if $\mu \le m$}\\
0 & \textrm{if $\mu > m$}
\end{array}\right.
\ee
(The primary is defined to be the more massive star.) Hillenbrand
\shortcite{H97} found the mass distribution of stars in the ONC to be
reasonably fitted by a Miller-Scalo mass function -- a semi log-normal,
falling from a maximum at around 0.1 \msol. Using this function for
$f(m)$, we evaluate the integral numerically, and obtain a result of
0.75 for $X$.

\begin{figure}
\epsfxsize=8.0truecm\epsfbox{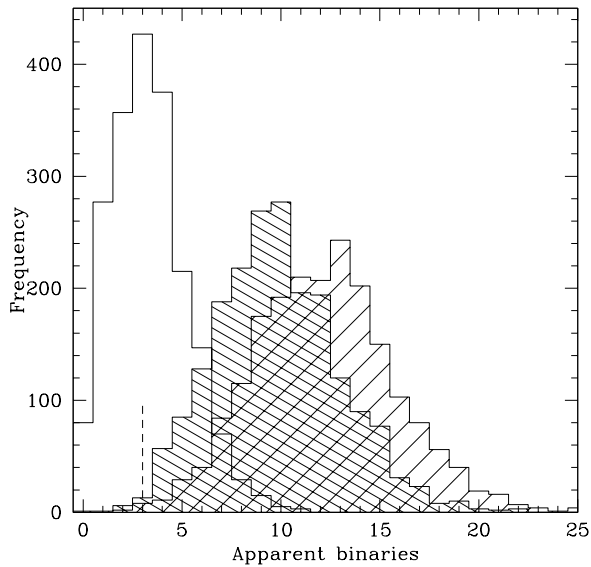}
\caption{Histogram showing the results of observing three ensembles of
simulated data, one having no binaries, and the other two (hatched) having
a binary population matching that found in the solar neighbourhood.
Each ensemble comprises 2000 clusters. The data with heavy hatching has
been corrected for a brighter detection limit in the cluster centre
than in the outer region, whereas the data with light hatching assumes
the fainter detection limit throughout (see text, \S 3). The dashed line
indicates the observational result of three apparent binaries for the
ONC\@.}
\label{F:results}
\end{figure}

\vspace{12pt} \noindent
With these corrections applied, each model cluster is then analysed for
apparent binaries, as was done with the JW data in the previous
section. The results for ensembles of 2000 clusters, plotted in Figure
\ref{F:results}, show that the
clusters with zero binary fraction, when seen in projection, are far
more likely to show three common proper motion binaries than the
clusters with a Duquennoy \& Mayor \shortcite{DM91} binary population.
Specifically, we find that in the former case, a result of three
apparent binaries occurs in 427 (21 per cent) of the 2000 clusters,
while for the Duquennoy \& Mayor clusters which have been corrected for
mass detection limits it occurs in just 13 (0.7 per cent) of them. A
result of three binaries is 49 centiles away from the median of the
latter distribution, which in a gaussian curve would correspond to a
distance of just over 2.3 times the standard deviation.

\section{Discussion}

Over the bulk of the cluster, Jones \& Walker found a roughly constant
three-dimensional velocity dispersion of about 4~$\mathrm{km\,s^{-1}}$,
increasing only slightly towards the core. They also found some
evidence for anisotropy in the velocity dispersion towards the outer
regions of the cluster, increasing in the radial direction -- the
direction towards and away from the cluster centre -- and decreasing in
the tangential direction. In terms of separation,
4~$\mathrm{km\,s^{-1}}$ corresponds to a hard/soft binary limit of
about 30~AU, much closer than the binaries we have considered here. Our
wide binaries are therefore well into the regime where one would expect
disruption through encounters with other cluster members.

The typical encounter time $\tau$ of a binary with another star in the
cluster can be crudely estimated as
\be
\tau = \frac{1}{n\sigma S}
\ee
where $n$ is the stellar density, $\sigma$ the velocity dispersion, and
$S$ the binary cross section (taken simply to be $\pi a^2$, with $a$
the semi-major axis, since we ignore the effect of gravitational
focusing for wide binaries). In the core, taking a stellar density of
$2 \times 10^4$~pc$^{-3}$, we expect a binary with $a = 1000$~AU to
have $\tau$ less than $10^5$ years. Further out, $\tau$ increases as
the density drops, surpassing $10^7$ years at about 1 pc from the
cluster centre (assuming an isothermal sphere distribution).

In her study of masses and ages in the ONC, Hillenbrand \shortcite{H97}
found the great majority of stars to be less than a million years old.
Since this is significantly shorter than the disruption timescale for
wide (1000--5000~AU) binaries in the outer regions of the cluster,
their apparent absence could be taken either as evidence that none were
formed in that separation range, or as an indication that all the stars
in the outer cluster have at some stage passed through a denser
environment, such as the core. Taking the core radius to be about 0.2
pc,\footnote{Note that this is not the same core radius as was used
previously in generating the model clusters. That value (0.04 pc)
simply parameterised the isothermal sphere distribution used to match
Jones \& Walker's data.} as found by Hillenbrand \& Hartmann
\shortcite{HHa98}, a star moving at 4~$\mathrm{km\,s^{-1}}$ will take
some $10^5$ years to pass through, so that even one visit would suffice
to disrupt most of these wide binaries. But the ONC is a young cluster,
and it is not clear that the majority of stars will have visited the
core in the time available.

Brandner \& K\"{o}hler \shortcite{BrK98} have examined the period
distribution of binaries in two subgroups of the Scorpius-Centaurus~OB
association, and found significantly more wide ($a > 200$ AU) binaries
in the subgroup containing fewer massive B type stars. It is plausible
that the binary period distribution in star-forming regions might be
determined at the time of formation, and be influenced by conditions
like the temperature and density of the cloud, as has been proposed
by Durisen \& Sterzik \shortcite{DuSt94}. However, it is
difficult to exclude the effects of subsequent dynamical evolution,
which can happen on a short timescale in these dense environments, and
which can account for a deficit of wide binaries independently of the
initial distribution. This effect has recently been investigated in
dynamical simulations of clusters in equilibrium, cold collapse, and
expansion by Kroupa, Petr \& McCaughrean \shortcite{KrPeMc99}. An
overview of dynamical processes in young clusters
is provided by Bonnell \& Kroupa \shortcite{BoKr99}.

In conclusion, we observe that about twenty per cent of the binaries in
Duquennoy \& Mayor's survey are wider than 1000 AU, so with a total
binary fraction of 0.6, they comprise around fifteen per cent of
G dwarf stars in the Galactic field. A similar figure can be assumed for
the M dwarfs in Fischer \& Marcy's \shortcite{FM92} survey.
If such systems do not come from
regions like Orion -- whether because they are never formed or because
they are dynamically disrupted -- then a significant number of field
stars must be formed elsewhere, possibly in cooler or less dense
environments.  This raises the more general question of what `mix' of
star forming regions is required to produce the field distribution we
see today, and demonstrates the utility of binary stars as a diagnostic
probe for such syntheses. Further collation of binary statistics in
different star forming regions, over the widest range of separations,
should continue to shed light on these questions.

\section{Acknowledgements}

We thank Ian Bonnell, Pavel Kroupa, and Melvyn Davies for valuable
help and discussions. A. Scally is grateful for the support of a
European Union Marie Curie Fellowship.

%\end{document}

\end{document}